\documentclass[11pt]{article}
\usepackage[]{amsmath,amssymb}
\usepackage{graphics,epsfig}

\begin{document}
\date{}
\title{A finite entanglement entropy and the c-theorem}
\author{H. Casini~\footnote{Email:casinih@ictp.trieste.it} and
 M. Huerta~\footnote{Email:huertam@icpt.trieste.it} \\
{\sl The Abdus Salam International Centre for Theoretical Physics, Strada
Costiera 11, 34100,}\\
{\sl Trieste, Italy }}
\maketitle

\begin{abstract}
The trace over the degrees of freedom located in a subset of the space
transforms the vacuum state into a mixed density matrix with non zero entropy.
This is usually called entanglement entropy, and it is known to be divergent in
quantum field theory (QFT).
However, it is possible to define a finite quantity $F(A,B)$ for two given different
subsets $A$ and $B$ which measures the degree of entanglement between their
respective degrees of freedom. We show that the function $F(A,B)$ is severely constrained
by the Poincar\'e symmetry and the mathematical properties of the entropy. In
particular, for one component sets in two dimensional conformal field theories its
 general form is
completely determined. Moreover, it allows to prove an alternative entropic version
of the c-theorem for $1+1$ dimensional QFT. We propose this well defined quantity as
the meaningfull entanglement entropy and comment on possible applications in QFT and the black hole
evaporation problem.
\end{abstract}


\section{Introduction}

The density matrix $|\Psi \rangle \langle \Psi |$ corresponding to the vacuum
state $|\Psi \rangle $ can be traced over the degrees of freedom located in
a subset of the space giving place to a mixed density matrix with non zero entropy.
This is usually called entanglement or geometric entropy. It was argued that this
quantity is deeply related to the entropy of black holes \cite{bom}. In
the black hole spacetime the causal structure provides a natural partition
of degrees of freedom, which corresponds to the asymptotic data at infinity
and at the horizon respectively. The quantum state relevant for asymptotic
observers is obtained by tracing the complete density matrix over the
invisible degrees of freedom on the horizon. But $|\Psi \rangle $ contains
correlations between the field modes inside and outside the black hole, and
in consequence, the partial trace leads to a mixed density matrix with non zero entropy.
Indeed, several different calculations in specific examples have shown that the geometric
entropy is proportional to the area in Minkowski space \cite{bom,sred,lw,ks}.
More recently the
entanglement entropy has been also a subject of much interest in the study
of condensed matter systems, specially in relation with the density matrix
renormalization group method \cite{vid,mas}.

More generally, the entanglement entropy of the vacuum state can be thought as a
special case of a more familiar quantity, the entropy $S(V)$ contained inside a
volume corresponding to a given region $V$ of the space. For gases and other
systems this is usually calculated by defining the system in question in
the bounded region, what requires the specification of boundary conditions.
However, these normally kill the correlations that the global state $\rho $
could present between degrees of freedom in $V$ and its complement $-V$. To
compute $S(V)$ without
artificially modifying the global state $\rho $ we need a partition of the
total Hilbert space $H$ as a tensor product of Hilbert spaces inside and
outside $V$, $H_{V}\otimes H_{-V}$. Then, the local density matrix is
defined by $\rho _{V}=Tr_{H_{-V}}\rho $ and its entropy is $S(V)=-Tr\rho
_{V}\log \rho _{V}$. The local Hilbert spaces $H_{V}$, density matrices
$\rho _{V}$ and entropies $S(V)$ must satisfy certain compatibility conditions
that simply express that they arise from a global state $\rho$ and Hilbert
space $H$ by partial tracing.
Given two non intersecting sets $A$ and $B$ we have that
\begin{equation}
H_{A\cup B}=H_{A}\otimes H_{B}\,,\;\;\;\;\;\;\;\;
\rho _{A}=Tr_{H_{B}}\rho _{A\cup B}\,.
\end{equation}

\begin{figure}
\centering
\leavevmode
\epsfysize=5cm
\epsfbox{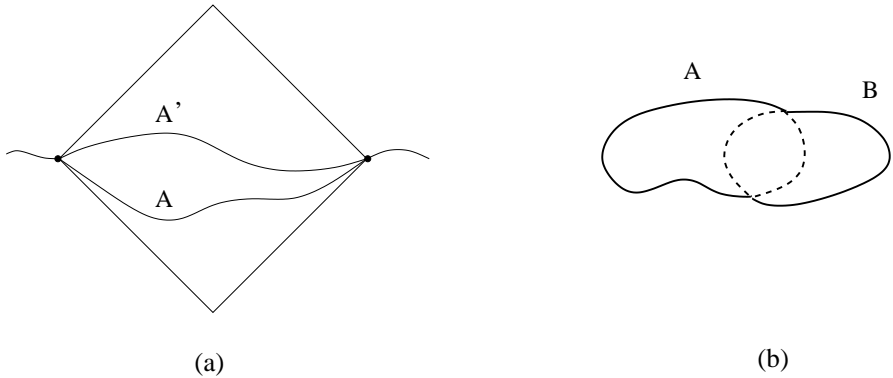}
\bigskip
\caption{(a) The spatial surfaces $A$ and $A^\prime$ have the same causal domain of dependence
 given by the diamond shaped set in this picture. The same entropy should correspond to $A$
 and $A^\prime$. Thus, we do not make distinctions between a
diamond set and any of its Cauchy surfaces. (b) Two spatial surfaces $A$ and $B$ included in the
same global Cauchy surface.
Writting $\text{area}(X)$ for the
$d-1$ dimensional volume of the boundary of a $d$ dimensional spatial set $X$ we have
$\text{area}(A)+\text{area}(B)=\text{area}(A\cap B)+\text{area}(A\cup B)$.}
\label{f1}
\end{figure}

These conditions allow us to use a very
important mathematical property of the quantum entropy which is called the
strong subadditive inequality (SSA) and which is usefull to relate the
entropies corresponding to the different subsystems involved in a tensor
product \cite{lr,we}.  In the present case it acquires the following
geometrical form \cite{we,rr,cas}
\begin{equation}
S(A)+S(B)\ge S(A\cap B)+S(A\cup B)\,,  \label{ssa}
\end{equation}
for any two sets $A$ and $B$.
In addition, the purity of the vacuum state gives the duality relation
\begin{equation}
S(A)=S(-A)\,.  \label{dual}
\end{equation}

In Minkowski space the independent degrees of freedom should be assigned to
subsets of general Cauchy surfaces and an entanglement entropy must
correspond to any of these subsets. Moreover, the causal structure and unitarity imply
that this entropy must be the same for different spatial sets having the same
causal domain of dependence (see fig.1(a)). It has been shown that the
combination of the SSA property, the positivity of the entropy, and the
Poincar\'e invariance of the vacuum state in Minkowski space
strongly constrain the function $S$ in any dimensions \cite{cas}.
The result is that if $S$ is
finite for at least one arbitrary set then it has to finite and  proportional to the boundary
area for any other set, plus a constant term:
\begin{equation}
S(A)=s\,\,\,\text{area}(A)+\alpha\,.  \label{area}
\end{equation}
Here area$(A)$ means the area of the boundary of the spatial surface $A$
and $s$ and $\alpha$ are positive constants. Thus, the area law for
the entanglement entropy  has its origin in very basic and fundamental
properties of relativity and quantum mechanics.  In $1+1$ dimensions the
equation corresponding to (\ref{area}) reads
\begin{equation}
S(A)=s\,\,m_A+\alpha\,,  \label{unouno}
\end{equation}
where $m_A$ is the number of connected components (or the number of boundary
points) of the one dimensional set $A$. The SSA relation was originally introduced
to show the existence of the mean entropy (the limit of entropy divided by volume
for large sets) in translational invariant systems \cite{we,rr}. We stress
that the presence
of additional boots symmetries makes the SSA relation a much more powerfull tool.
For example, the equation (\ref{unouno}) tells that the entropy of a single component
set in $1+1$ dimensions is a constant (if it is finite), while the SSA
relation combined with translational symmetry alone
only implies that this is a concave function of the length. For different types of sets
or more dimensions the contrast is much greater.

The equations (\ref{area}) and (\ref{unouno}) apply for finite $S$. However, all specific
calculations, numerical and analytical, have shown that the entanglement entropy is
ultraviolet divergent, even for free fields. This divergences can not be
softened by adding different matter fields since the entropy is a positive quantity.
This indicates that the
entanglement entropy may not have a finite covariant meaning in quantum
field theory (QFT).

The calculations of $S$ present in the literature have shown that
the divergent terms are proportional to the
area, which is also suggested by the theorem leading to eq. (\ref{area}) \cite{cas}.
Motivated by
this fact, we introduce a new quantity constructed out of the function $S$ which is
however free from divergences.
Let $A$ and $B$ be two
spatial sets in Minkowski space which belong to the same Cauchy
surface~\footnote{ Constructions with two sets
 where also used in a related context to calculate quantum fluctuations of operators attached to
the local regions \cite{y}.}
 (see fig.1(b)). The symmetric function
\begin{equation}
F(A,B)=S(A)+S(B)-S(A\cap B)-S(A\cup B) \label{efe}
\end{equation}
 is finite, since the divergences cancel due to the relation
\begin{equation}
\text{area}(A)+\text{area}(B)=\text{area}(A\cup B)+\text{area}(A\cap B)\,.\label{area1}
\end{equation}
It is also positive because of strong subadditivity and satisfies the duality
\begin{equation}
F(A,B)=F(-A,-B)\,. \label{dualf}
\end{equation}

Besides, $F(A,B)$ has a very suggestive property, not shared by $S$. When $A$ and $B$
are non intersecting sets $F(A,B)$ is monotonically increasing,
\begin{equation}
F(A,B)\le F(A,C) \,\,\,\, ,\,\,\,\,\ B \subseteq C \,\,\, , \,\,\,  C \cap A=0 \,.
\end{equation}

 From all this, we are lead to propose $F$ as a well defined measure of
the degree of entanglement in QFT. In fact, for two non intersecting sets $A$ and $B$ it
can be written as
\begin{eqnarray}
F(A,B) &=& S(A)+S(B)-S(A\cup B) \nonumber \\
&=& Tr(\rho _{A\cup B}\log \rho _{A\cup B})-Tr(\rho
_{A\cup B}\log (\rho _{A}\otimes \rho _{B})) \, = S(\rho _{A\cup B}\mid \rho _{A}\otimes \rho _{B}),
\label{caso}
\end{eqnarray}
where $S(\rho _{1}\mid \rho _{2})=Tr(\rho _{1}(\log (\rho _{1})-\log (\rho _{2}))
$ is known as the relative entropy for two states $\rho _{1}$ and $\rho _{2}$
acting on the same Hilbert space. This later can be thought as a measure of
 the statistical distance between
 a pair of states. In the context
of quantum information theory, the particular relative entropy in (\ref{caso})
is called the mutual information
between the states $\rho_A$ and $\rho_B$ in the composite quantum system, and
 it is used as
a measure of the amount of information they have in common \cite{libro}.
The function $F(A,B)$ for
intersecting sets can be defined in terms of the relative entropy for non
intersecting ones with the use of additional auxiliary sets.

The connection between $F$ and the relative entropy is also satisfying from the
mathematical point of view.
 The origin for the divergence of the standard entanglement entropy can be
traced to the impossibility of expressing the Hilbert space $H$ as a tensor
product of $H_V \otimes H_{-V}$ in the relativistic case.
This partition is unambiguous if the system
is defined on a lattice, the local
Hilbert spaces being generated by the local degrees of freedom (for example
the spin-like operators in a given region). However, in a relativistic QFT
the axiomatic investigations have shown that although there is a well
defined notion of local algebras of operators in a volume $V$ (type III Von
Neumann algebras in general) there is no Lorentz invariant partition of the
total Hilbert space into a tensor product $H_{V}\otimes H_{-V}$ (see \cite
{haag,hal}). One heuristic reason for this can be seen
considering the theory of a free scalar field $\phi $. In the
classical theory the phase space attached to a bounded region $V$ is given
by initial data $\phi (x)$ and $\dot{\phi}(x)$ vanishing outside $V$. In the
quantum case however, only positive energy solutions of the wave equation
 appear in the annihilation part of the operator $\phi (x)$.
In momentum space the time derivative is then
$\epsilon_p=-i\sqrt{p^{2}+m^{2}}$.
Now, in real space, the operator given by $\epsilon_p$
is antilocal, what means that it transforms any function vanishing outside
an open set $V$ in a function which does not vanish identically in any open
subset outside $V$ \cite{hal}. This implies that one can not have
$\langle 0\vert \phi(x)\vert \psi \rangle $ and $\langle 0 \vert \dot{\phi}(x) \vert \psi \rangle $
vanishing simultaneously in
the same region of space, where $\vert \psi \rangle$ is any one particle state.
In general the moral is that in QFT
there is no covariant meaning for the localization of states, and what can
be unambiguously localized are the field operator algebras. In this
sense it is remarkable that even if a notion of entropy for a state in a
given algebra may not exist
(type III algebras do not contain a trace), the relative entropy of two
states for the same Von Neumann algebra has a well defined mathematical
meaning \cite{ara}.

\section{Entanglement entropy in $1+1$ dimensions}

The equation (\ref{efe}) implies the following relation between the finite $F$ functions
\begin{eqnarray}
F(A,B)+F(A,C)+F(A\cup B,A\cup C)&+&F(A\cap C,A\cap B) \nonumber\\
&=&F(B,C)+F(A,B\cup C)+F(A,B\cap C)\,,\label{equ1}
\end{eqnarray}
for any $A$, $B$ and $C$. In a given regularization scheme, as $S$ goes to infinity these relations
between finite quantities are maintained. The most
general form of a function that solves (\ref{equ1}) in $1+1$ dimensions can be found by a long but
straightforward calculation that we are not presenting here. It follows that
\begin{equation}
F(A,B)=G(A)+G(B)-G(A\cap B)-G(A\cup B)\,. \label{g}
\end{equation}
Evidently, this equation has the same form as (\ref{efe}) but now
the one set function $G$ is perfectly finite. Since the meaningfull quantity is $F$,
the function $G(A)$ is defined up to an arbitrary term proportional to the number $m_A$
of connected components in $A$. Thus, for the one component sets there is a "gauge"
symmetry associated to an additive constant. The positivity of $F$ implies strong subadditivity
for $G$,
\begin{equation}
G(A)+G(B)\ge G(A\cap B)+G(A\cup B)\,,  \label{ssag}
\end{equation}
and from (\ref{dualf}) we also have $G(A)=G(-A)$.

The crucial difference between $G$ and an entropy is that it can attain negative values.
This must be so if $G$ has some dependence on the size of the sets, because otherwise, as we have
already mentioned, eq. (\ref{unouno}) would apply for $G$. Moreover,
it must attain negative values with
arbitrarily big modulus. If not, it could be converted into a
positive function by adding a suitable
term proportional to $m_A$.

\subsection{Entanglement entropy in Conformal Field Theory}

 For the special case of a conformal field theory (CFT) and when
the sets $A$ and $B$ are one component
and intersecting, the eq. (\ref{g}) allows us to obtain the general form of $F(A,B)$.
We see from figure 2(a) that only four points
determine the position of the diamonds corresponding to $A$ and $B$. As $F(A,B)$ must be
invariant under global conformal transformations, general results ensure that it must
be a function of the cross ratios
\begin{equation}
F(A,B)\equiv F(\eta_u,\eta_v)\,.\label{equiv}
\end{equation}
with
\begin{equation}
\eta_u=\frac{u_{23}\,u_{14}}{u_{13}\,u_{24}}\,\,\,\,\,\,\,,\,\,\,\,\,\,\,\eta_v=
\frac{v_{23}\,v_{14}}{v_{13}\,v_{24}}\,,
\end{equation}
where we have written $u_{ij}=u_i-u_j$ and $v_{ij}=v_i-v_j$ and
the $(u_i,v_i)$ are the null coordinates of the point $x_i$.
On the other hand, for one component sets, $G$ can be written as a function of the
length of the diamond base
$r_{ij}=\sqrt{(x_{i}-x_{j})^2}=\sqrt{u_{ij}v_{ij}}$.
 Using this and  equations (\ref{g}) and (\ref{equiv})
we get the general form for $G$ and $F$
\begin{eqnarray}
G(A)&=& k \log(r_A)+\beta\,.\label{ga}\\
F(A,B)&=& k \log\left(\frac{r_A r_B}{r_{A\cap B} r_{A\cup B}}     \right)\,,\label{forma}
\end{eqnarray}
where $k$ and $\beta$ are constants, with $k\ge 0$.

The conformal symmetry provides a simple argument in order to explain why $F$ is
finite while $S$ is not. Both quantities are supposed to be
invariant under global conformal transformations which leave the vacuum invariant,
but while $F$ is a
function of four points only two points determine
the argument of $S$ for one component sets, and there is no possibility
of forming a cross ratio with two points. Thus
$S$ must be either infinite or a constant.

Several authors have calculated the entanglement entropy $S$ in $1+1$ conformal field
theories with different
regularization prescriptions, both numerically and analytically \cite{sred,ks,vid,hlw}.
The result is $S=(1/6)(c+\bar{c}) \log(r_A/\epsilon)$, where $\epsilon$ is an ultraviolet
cutoff, and $c$ and $\bar{c}$ are the holomorphic and antiholomorphic central charges.
This formula can not be used for $l_A\le \epsilon$ that gives $S\le 0$.
According to equation (\ref{efe}), our results (\ref{ga}) and (\ref{forma}),
obtained from very general
considerations, are perfectly consistent with this expression for $S$, identifying $k$ with
$(c+\bar{c})/6$.
Thus, $G$ can be interpreted as a regularized entropy, $G(A)=k\log(r_A \mu)$,
with $\mu$ a renormalization scale. The fact that $G$ is negative in some range causes
no problem since
its physical meaning is only up to a constant term.

From eq. (\ref{forma}) we see that the divergences appear in $F$ reduced to
 certain special configurations of the
diamonds.
This is when $r_{A\cap B}$ goes to zero, which is not surprising, since in
that limit the two diamonds
 separate each other and $F$ turns into a different function,
the one for non intersecting one component diamonds we discuss below.

\begin{figure}
\centering
\leavevmode
\epsfysize=5cm
\epsfbox{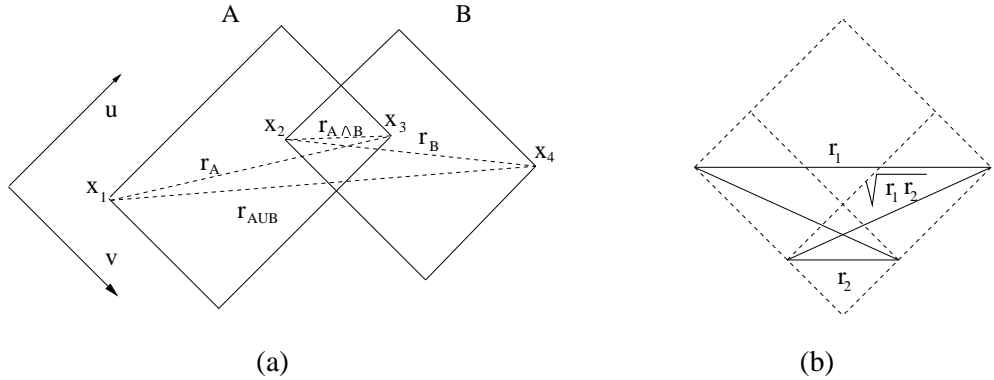}
\bigskip
\caption{(a) Two intersecting one component sets $A$ and $B$ whose straight Cauchy surfaces
have sizes $r_A$ and $r_B$ respectively. The diamonds formed by the intersection and union
(followed by causal completion) of $A$ and $B$ have sizes
$r_{A\cap B}$ and $r_{A\cup B}$. This configuration of sets is uniquely determined by the
position of the points $x_1, x_2, x_3$ and $x_4$ forming the spatial corners of the
diamonds. (b) Configuration of two intersecting sets of size $\sqrt{r_1 r_2}$ from which eq.
(\ref{relativista}) can be
obtained by means of the SSA inequality.}
\label{f2}
\end{figure}

\subsection{An entropic c-theorem}

Now we look for the general form of $G(A)\equiv G(r_A)$ for one component sets
in $1~+~1$ dimensional theories.
Consider two one component sets $A$ and $B$ of the same size $r_1$, intersecting in a segment
of size $r_1-r_2$
with $r_2 \leq r_1$. It follows from (\ref{ssag}) applied to $A$ and $-B$ that
\begin{equation}
G(r_1)\geq G(r_2)\,\,\,\,\,\,\,,\,\,\,\,\,\,\, r_1\ge r_2\,.
\end{equation}
Thus, $G(r)$ is increasing and $G^\prime\ge 0$. The SSA relation for intersecting
diamonds disposed as shown in the figure 2(b) reads
\begin{equation}
G(\sqrt{r_1 \, r_2})\ge \frac{1}{2} (G(r_1)+G(r_2))\,. \label{relativista}
\end{equation}
It can be shown that all other inequalities
 coming from SSA for two intersecting one component diamonds are less constraining
than this relation. The meaning of (\ref{relativista}) becomes more clear defining the function
\begin{equation}
\bar{G}(a)=G(e^a)\,,
\end{equation}
such that the domain of $\bar{G}$ is the whole real line. Equation (\ref{relativista}) becomes
\begin{equation}
\bar{G}(\frac{1}{2}(a+b))\ge \frac{1}{2}(\bar{G}(a)+\bar{G}(b))\,.
\end{equation}
This relation is called weak concavity, and it is known to be a property of the entropy for translational invariant
systems in the original length variable. Here we have it in the logarithmic variables. Using repeatedly this
equation we can prove concavity for $\bar{G}$, that is
\begin{equation}
\bar{G}(\lambda a+(1-\lambda) b)\ge \lambda \bar{G}(a)+ (1-\lambda) \bar{G}(b) \,,\label{concave}
\end{equation}
for any $\lambda\in [0,1]$.
The concavity of $\bar{G}$ can be more simply written $\bar{G}^{\prime\prime}\le 0$.
Introducing a new function
$C(r)\equiv r\,G^\prime (r)$ we summarize all the constrains on
the original function $G$ as
\begin{eqnarray}
C(r)\equiv r\,G^\prime (r)\ge 0\,,\label{c1}\\
C^\prime(r)= G^\prime(r)+ r \,G^{\prime \prime}(r)\le 0\,.\label{c2}
\end{eqnarray}

According to the equations (\ref{c1}) and (\ref{c2}) the dimensionless function $C$
must be positive and decreasing with $r$. Since it depends on $G^{\prime}$ it is also a
physically
meaningfull quantity independent of the ultraviolet cutoff. 
Thus, the renormalization group equation for $C$ reads
\begin{equation}
\tau\frac{\partial}{\partial \tau} C=-\sum_i \beta_i({g}) \frac{\partial}{\partial g_i} C \,,
\label{rg1}
\end{equation}
where $\tau$ is the substraction point, the $g_i$ are the different
dimensionless coupling constants and $\beta_i({g})$ the corresponding beta functions.
This equation expresses that the total derivative of $C$ with respect to 
$\tau$ is zero, what follows from the fact that $C$ is an observable quantity in a renormalizable theory. 
Note that in consequence, the anomalous dimension term is not present in (\ref{rg1}).
By dimensional analysis $C$ has to be a funtion of $(\tau\,r)$ and the dimensionless coupling 
constants, then we have 
\begin{equation}
\left( r\frac{\partial}{\partial r}- \tau\frac{\partial}{\partial \tau} \right) C\,=0\,.
\label{ren}
\end{equation}
As $C$ is decreasing with $r$ we see from
eq. (\ref{rg1}) and (\ref{ren}) that $C(r_0,{g})$ for fixed $r_0$ is a
function of the coupling constants which decreases along
the renormalization group trajectory. Moreover, at the fixed points, corresponding
 to conformal field theories, eq. (\ref{ga}) implies that $C(r_0,{g})$ is a constant
$k=(c+\bar c)/6$, stationary under the renormalization group flow. This is
 in fact the
statement of the Zamolodchikov theorem \cite{zam}.

Remarkably, this theorem follows here from the properties of the entropy and the
Poincar\'e symmetry
without resorting to quantum fields. Also, the unitarity and causality of the theory are implicitly 
used in the fact that the entropy of a given set does not depend on the particular choice of
the Cauchy surface, what is crucial for the arguments that lead to eq. (\ref{relativista}) (see 
figure 2(b)). Moreover, the SSA property requires a self-adjoint and positive density matrix which 
is also related to unitarity (however, it also works for classical probability 
distributions)\cite{lr,we}. 

The $C$ charge in eq. (\ref{c1}) has an interesting physical meaning: it measures the
variation of the geometric
entropy with the size of the set whose degrees of freedom are traced over. For example,
for massive fields,
the entropy saturates at a given radius (see for example \cite{vid}),
what leads to the vanishing of $C$ for large r.
This is in accord with the renormalization group flow interpretation of the
running of $C$ given by eq. (\ref{ren})
 since at the
 infrared no available degrees of freedom remain in the massive case.

In relation with the massive $C$ function we make the following remark that gives
 additional support to the connection of the entropy $F$ with
 the algebraic approach to QFT. Numerically we find that
the $C$
function (\ref{c1}) for a free scalar field of mass $M$ decreases exponentially,
 $C(r)\sim e^{-\sqrt{3} M r}$, for large $r$,
 what also happens for the Zamolodchikov $c$ function. Adding independent scalar fields
 we have a contribution to $C(r)$ due to the sum of the exponential tails which is of the form
 $\int dM \sigma (M) e^{-\sqrt{3} M r}$, where $\sigma (M)$ is the density of fields
per unit mass. We can choose an exponentially increasing density
$\sigma (M)\sim e^{\frac{M}{T_H}}$ in such a way that the theory has a maximal temperature
 $T_H$ (Hagedorn temperature).  In this case the divergence of $G$ does not take place at $r=0$
but at a finite distance $r\sim T_H^{-1}$. This coincides with the results of \cite{buc} which
 tell that a consistent tensor product structure can not be given to the states on the
algebras of two non intersecting sets when their separation distance is less than some constant times
$T_H^{-1}$.

Some relations between the geometric entropy and the renormalization group flow were already
suggested in the literature based on
  numerical results \cite{vid}. Moreover, interesting entropic c-theorems were introduced
  in \cite{gai}. These also rely on the properties of the entropy.  However, they do not involve
   the entanglement entropy between the degrees of freedom
   in different regions of the space but the relative entropy between the vacuum states corresponding
    to different values of the theory couplings.

The renormalization group flow of the Zamolodchikov $c$ function is given in terms
of the energy momentum tensor correlators. It is likely that the energy momentum tensor
is also involved in the running of the entropic function $C$ proposed in this paper. This
would be very interesting, specially taking into account that one of the originals
aims of the investigations about the geometric entropy was the underestanding of
the relation between
entropy  and gravity.

\subsection{The two component set function}
The equations (\ref{c1}) and (\ref{c2}) give the most general form of the function
$F(A,B)$ in $1+1$ dimensions for intersecting one component sets $A$ and $B$.
 Of course, these leave a lot of freedom, but this must be so, since
$F$ contains information on the mass spectrum of the theory. The analysis can be continued
considering the multicomponent set functions,
that is, the case in which the sets $A$ and $B$ are multicomponent or non intersecting.
 Then, the function $F$ and the relevant inequalities involve the function $G$ evaluated on
multicomponent sets. The general analysis
is outside of the scope of this letter. Here we show only some constrains for the simple case
 of two non intersecting one component sets $A$ and
$B$ lying in a single spatial line in a CFT. These sets are as before determined by the position of
their four edge points, what implies due to global conformal invariance that $F$ is a
function of the cross ratio
\begin{equation}
F(A,B)=F(\eta)=k\log(r_A)+k\log(r_B)-G(A\cup B)\,,\label{vein}
\end{equation}
with
\begin{equation}
\eta=\frac{r_A r_B}{(r_A+r_C)(r_C+r_B)}\,,
\end{equation}
where $r_A$ and $r_B$ are the sizes of $A$ and $B$ and $r_C$ is their separating distance.
 We can rewrite eq. (\ref{vein}) as
\begin{equation}
G(A\cup B)=k\, \log\left( \frac{r_A r_B r_C (r_A+r_B+r_C)}{(r_A+r_C)(r_C+r_B)}\right)
-U(\eta) \,,
\label{gab}
\end{equation}
with $U(\eta)=F(\eta)+k\log (1-\eta)$.
The duality (\ref{dualf}) leads to
\begin{equation}
U(\eta)=U(1-\eta)\,.
\end{equation}
Besides, from the SSA relation between $A\cup B$ and another set $C$ with one or two
components it follows that
\begin{eqnarray}
U(\eta)&\ge& k \log(\eta)\,,\label{ineq1}\\
U^\prime (\eta) &\le& \frac{k}{\eta}\,,\label{ineq2}\\
U^{\prime\prime} (\eta)&\ge& -\frac{U^\prime (\eta)}{\eta}-\frac{k}{\eta(1-\eta)^2}\,.\label{ineq3}
\end{eqnarray}
These inequalities highly constrain the function $U$, and in particular show that it has
to be bounded in its
domain $\eta\in [0,1]$. Thus, the only singular points in $G(A\cup B)$ come from the
first term in the right hand side of (\ref{gab}).
 In particular in the limit $B\rightarrow -A$, $r_C=\epsilon\rightarrow 0$ we get
$F\rightarrow 2k \log (r_A /\epsilon)$. From eq. (\ref{efe}) this gives
\begin{equation}
S(r_A)\simeq k \log(l_A/\epsilon)\,,
\end{equation}
consistent with previous calculations \cite{hlw}.

We have checked numerically the formula (\ref{gab}) for a Majorana and a Dirac
fermion ($c=1/2$ and $c=1$ respectively) using the methods of \cite{vid,pes}. The result is
$U\equiv 0$ and $k=1/6$ and $k=1/3$ respectively. On the other hand,
we have also considered the theory of a
free massless boson ($c=1$). The existence of a zero mode provides
an additional infrared divergence to the entanglement entropy which is also present in $F$
(this happens for all theories with non compact global symmetry groups).
 To regularize it we have introduced a small mass $M$ and attached the theory to a cylinder of
circumference $L$ such that $ML\ll 1$. In this case the numerical
result is
\begin{equation}
U(\eta_{\text{cyl}})=\frac{1}{9} \log(\eta_{\text{cyl}} (1-\eta_{\text{cyl}}))
-\frac{1}{2} \log (M L), \label{for}
\end{equation}
where
\begin{equation}
\eta_{\text{cyl}}=\frac{\sin(\frac{\pi r_A}{L} )\sin(\frac{\pi r_B}{L})}
{\sin(\frac{\pi (r_A+r_C)}{L})\sin(\frac{\pi (r_B+r_C)}{L})}\,,
\end{equation}
is the expression of $\eta$ in the cylinder. With $r_A\,,r_B\,,r_C\,\ll L$ one recovers
$\eta_{\text{cyl}}\rightarrow \eta$.
The formula (\ref{for}) satisfies the inequalities (\ref{ineq1}-\ref{ineq3}) with $k=1/3$, except
when $\eta\lesssim (ML)^{(9/2)}$ or $(1-\eta)\lesssim (ML)^{(9/2)}$ where the mass enters into
the game
and equation (\ref{for})
and the conformal symmetry
do not hold any more. This is why (\ref{for})
does not seem to lead to a bounded $U$.
The point we want to make here is that the function $U$ has the ability to distinguish
between different models with the same central charge ($c\ge 1$).

\section{Final remarks}
One could wonder if the whole set of entanglement entropies are in one to one correspondence
with
the different QFT. In other words, is it possible to uniquely define  a QFT by the set
of its entanglement entropies?. This is likely, since the local algebras of all QFT are the
same mathematical objects, independent of the theory and the space-time dimension \cite{haag}.
Thus, the way they are
entangled to each other must play a central role. If one accepts
this, a second question naturally appears: is it the only requisite of a consistent set of
entropy functions to obey the SSA relations and the space-time symmetries?.

\begin{figure}
\centering
\leavevmode
\epsfysize=5cm
\epsfbox{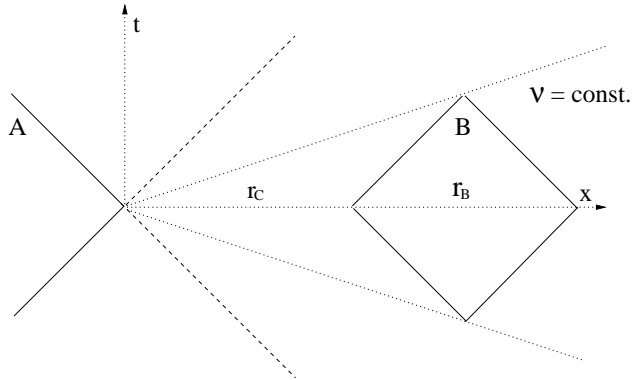}
\bigskip
\caption{The Rindler wedge $A$ and an exterior diamond set $B$. The curves of constant $\nu$ are
 lines through the origin.}
\label{f3}
\end{figure}

Further interest to these questions is given by the fact that
 the entanglement entropy $F$ has a nice property suited
to study dimensional reduction.
Take for
 example the sets of the form $A\times [x]$ in $d+1$ spacial dimensions, where $[x]$
is an interval of length $x$ and $A$ is any $d$ dimensional set. Then the function
\begin{equation}
\bar{F}(A,B)=\lim_{x\rightarrow \infty} \frac{F(A\times [x],B\times [x])}{x}
\end{equation}
is well defined and must satisfy all requirements of an $F$ function in $d$
dimensions \cite{cas}. Generalizations of the results of this paper to
higher dimensions using this construction are currently under investigation.

Finally, we suggest an interpretation of $F$ in the black hole evaporation problem.
 Take $A$ as the Rindler wedge and let $B$ be a diamond outside $A$ as shown in fig. 3.
 Then $F(A,B)$ can be interpreted, at least in certain limit, as
the amount of entropy from the Unruh thermal radiation
 that an observer can detect in an
experiment running for a finite period of time corresponding to the set $B$
 (see related ideas in \cite{ks,km}).
This period measured in the
boost parameter $\nu=\text{arcth} (t/x)$ is
\begin{equation}
\Delta \nu=\log\left(\frac{r_B+r_C}{r_C}\right)\,.
\end{equation}
Now, the equations (\ref{vein}) and (\ref{gab}) for $r_A\rightarrow \infty$ give
\begin{equation}
F(A,B)=k \Delta \nu + U\left( \frac{r_B}{r_B+r_C}\right)\,.
\end{equation}
Thus, as $U$ is bounded, for big enough $\Delta \nu$ this gives
$F(A,B)\simeq k \Delta \nu$. We also have $\Delta \nu=a \Delta \tau$, where $\tau$ is
the proper time meassured
by an accelerated
observer with constant acceleration $a$.
This formula tells that the entropy is proportional to the diamond
time $\Delta \tau$ times a constant flux of
entropy $k a$ due to a constant flux of Unruh radiation. Indeed, the Unruh temperature
 is $T=a/(2\pi)$ and in the high temperature limit (adequate to the limit taken above)
 the flux of entropy per
unit time in a CFT has a universal
expression given by $\Delta s/\Delta \tau =\pi \frac{c+\bar{c}}{3}  T$ \cite{car}.
 Therefore, the identification of $F$ with the total entropy flux during the interval
 $\Delta \tau$ lead us again to the equation $k=\frac{c+\bar{c}}{6}$.
In the black hole case $\Delta \nu$ should be interpreted as $\Delta \tau/(4M_{BH})$, where $\tau$
is the time as measured by the asymptotic observers and $M_{BH}$ is the black hole mass.
Thus, $F$ would be
proportional to the asymptotic time, corresponding to a constant flux of Hawking radiation.
For a real black hole the back-reaction
must be included. In four dimensions the black hole lifetime is $\Delta \tau\sim M_{BH}^3$, giving
place to a maximal observable entropy $F(A,B)\sim M_{BH}^2$, which is proportional to the horizon
 area.

\section{Acknowledgments}
H.C. thanks the invitation to visit the ICTP where this work has been done.

\vspace{0.3cm}

{\sl Note added}.- After the submission of this paper to the arXiv data-base, the reference \cite{cwc} appeared, 
showing a calculation of the multi-set CFT entropy functions. Their general results are in agreement with 
the particular cases considered in Sections 2.A and 2.C.    

\vspace{0.3cm}
{\sl Erratum:}

\noindent 1.- Equation (\ref{area1}) eliminates the divergences in $F(A,B)$ only in two dimensions. In more dimensions additional subleading terms are present and $F(A,B)$ does not converge for generic $A$ and $B$. 

\noindent 2.- In Section 2.B the long distance behaviour  $C(r)\sim e^{-\sqrt{3}Mr}$ should read $C(r)\sim e^{-2Mr}$.

\noindent 3.- In Section 3 the large $\Delta \nu$ limit is the low (rather than high) temperature limit.  There is a factor two in $\Delta s/\Delta \tau $ which should read $\pi \frac{c+\bar{c}}{6} T$. This means that $F$ represents {\sl twice} the total entropy flux during the time $\Delta \tau$.

\end{document}